\documentclass[aps,prd,
 ]{revtex4}
\usepackage{graphicx}
\usepackage{color}
\usepackage{slashed}
\usepackage{amsmath}

\newcommand\bef{\begin{figure}}
\newcommand\eef[1]{\label{fg:#1}\end{figure}}
\newcommand\beq{\begin{equation}}
\newcommand\eeq[1]{\label{#1}\end{equation}}
\newcommand\beqa{\begin{eqnarray}}
\newcommand\eeqa[1]{\label{#1}\end{eqnarray}}
\newcommand\bet{\begin{table}}
\newcommand\eet[1]{\label{tb:#1}\end{table}}

\newcommand\fgn[1]{Figure \ref{fg:#1}}
\newcommand\eqn[1]{eq.\ (\ref{#1})}

\newcommand\etc{{\sl etc.\/}}
\newcommand\ie{{\sl i.e.\/}}

\newcommand\tco{T_{co}}
\newcommand\tfo{T_{fo}}
\newcommand\Neq{n^{{\rm eq}}}
\newcommand\npi{n_\pi}
\newcommand\nk{n_{\scriptscriptstyle K}}
\newcommand\nkb{n_{\overline{\scriptscriptstyle K}}}
\newcommand\neta{n_\eta}

\begin{document}

\title{Chiral symmetry breaking and chemical equilibrium in a heavy-ion collisions}
\author{Sourendu\ \surname{Gupta}}
\email{sgupta@theory.tifr.res.in}
\affiliation{Department of Theoretical Physics, Tata Institute of Fundamental
         Research,\\ Homi Bhabha Road, Mumbai 400005, India.}
\author{Jajati\ K.\ \surname{Nayak}}
\email{jajati-quark@vecc.gov.in}
\affiliation{Variable Energy Cyclotron Centre, Kolkata 70064, India.}
\author{Sushant\ K.\ \surname{Singh}}
\email{sushantsk@vecc.gov.in}
\affiliation{Variable Energy Cyclotron Centre, Kolkata 70064, India, and\\
 Homi Bhabha National Institute,\\ Training School Complex, Anushakti Nagar,
 Mumbai 400085, India.}

\begin{abstract}
We examine the thermalization of an ensemble of the octet of pseudoscalar
mesons, in the isospin symmetric limit, whose interactions are
constrained through chiral symmetry, unitarity, and measurements. The
reaction amplitudes generate all resonances up to masses of about 2 GeV,
with twelve input parameters, namely $f_\pi$, three masses, and eight
low energy constants (LECs) of chiral perturbation theory.  In linear
response theory, we find that matter takes an extremely long time to
thermalize. These long relaxation times are directly related to the
fact that these mesons are pseudo-Goldstone bosons of chiral symmetry
breaking. This result indicates that fireballs created with zero baryon
number in heavy-ion collisions will drop out of chemical equilibrium
once they enter the chiral symmetry broken phase.\\

TIFR/TH/20-40
\end{abstract}

\maketitle

\goodbreak\section{Introduction\label{sec:intro}}

In heavy-ion collisions, as in all collider experiments, the main
observables are the particles in the final state and their momenta,
everything else of interest is constructed from these. The simplest
of observables are the abundances of hadrons in the final state. In
heavy-ion collisions, these yields are well explained by an ideal gas of
hadronic resonances at a freeze out temperature of $T_{fo}=158.4 \pm1.4$
MeV \cite{Andronic:2017pug} when the nucleon-nucleon center of mass
collision energies, $\sqrt S$, is larger than 20 GeV. Furthermore,
for $\sqrt S>100$ GeV, the flavour chemical potentials are small.
Many variants of such models have been examined \cite{variants}, and
they are in reasonable agreement with this result.

A coincidence arises from QCD with its approximate chiral symmetry.
In the limit of exact chiral symmetry, there would have been
a critical point at finite temperature, and in thermodynamic
equilibrium.  Since the symmetry is approximate, instead there is a
broad cross over \cite{Aoki:2009sc,Bazavov:2018mes} with a peak in the
chiral susceptibility at $\tco=156.5\pm1.5$ MeV. This has led to the
identification of $\tfo$ with $\tco$ \cite{Andronic:2017pug}.  At present
this is the only point of contact between heavy-ion collisions and the
broken chiral symmetry of hadronic physics.

Right from the early days of heavy-ion collisions there have been efforts
to build a complete dynamical description of the whole history of the
fireball in terms of a transport theory. These transport computations
assume that the fireball is made up of an interacting system of quarks and
gluons in the very initial stages, trace their interactions and usually
find that it approaches equilibrium, cools, and turns into an interacting
system of hadrons. In this widely accepted view of the fireball, the
number and momentum distribution of the final state hadrons is a result
of strong interaction dynamics. $\tfo$ is the freezeout temperature, \ie,
the point at which the expansion rate of the fireball matches the rate
of interactions, and chemistry, and eventually momenta, can no longer
be kept in thermal equilibrium through interactions.

Codes which implement these approaches, like URQMD \cite{Bleicher:1999xi}
and AMPT \cite{Lin:2004en}, incorporate a lot of known physics of strong
interactions and try to predict the complete course of a collision. The
hadronic interactions which are included in these approaches use many
two-to-two hadron processes.  A large fraction of these are not measured
yet, and have to be constrained by various model considerations. We partly
follow this approach in the sense that we examine transport theory with
the approximation of two-to-two interactions between hadrons. However,
unlike those codes, it is not our aim to model the full course of every
collision across a large range of $\sqrt S$.

Instead we ask a more limited question; namely, what would be the
relaxation time in a hadron gas pushed slightly out of chemical
equilibrium and allowed to relax back to equilibrium. The first step is
to decide which hadrons should be included. Certainly pions, which are
the lightest of hadrons, need to be accounted for. It turns out that
the lightest baryon, \ie, the proton, has an equilibrium number density
which is almost two orders of magnitude smaller at $T_{fo}$. Since
pion-proton and pion-pion cross sections are roughly comparable, within
the accuracy of a few percent in the relaxation rates, one may neglect
the baryon contribution in this computation. In this first study we will
do this. Inclusion of strangeness requires us to add kaons, the lightest
strange particle, to the mixture. Then the SU(3) flavour structure would
require us to add the $\eta$. So, the model contains the full octet of
the light pseudoscalar mesons. These are the pseudo-Goldstone bosons of
approximate chiral symmetry in QCD.

Even though it is approximate, chiral symmetry is predictive because
it strongly constrains the low-energy interactions of the pseudoscalar
mesons which are the pseudo-Goldstone bosons of this symmetry breaking
\cite{Goldstone:1962es, Weinberg:1978kz, Gasser:1983yg}. Such a theory
is known to be very good at predicting many properties of the lightest
pseudoscalar mesons including decay constants and reaction cross sections
\cite{Gasser:1983yg, Amoros:2000mc, GomezNicola:2001as}.  In this work
we use the unitarized cross sections of \cite{GomezNicola:2001as}.

It turns out that they reproduce the full resonance spectrum, at least
up to masses of 2 GeV. So, using the pseudoscalar octet with unitarized
amplitudes from chiral perturbation theory seems to capture a large
part of the mesonic physics that goes into the resonance gas model,
while also giving enough information to make a start on transport
computations. Among other pleasant aspects of the computation is that
it requires a very small number of input parameters, namely $m_\pi$,
$m_K$, $m_\eta$, $f_\pi$, and eight low energy constants of the chiral
effective theory (EFT) model. The amplitudes are unitary and contain
no further UV cutoffs. The eight LECs are obtained from other hadronic
observables. We will keep track of the error bands on all of them.

Some limitations of this computation are clear enough. The neglect of
baryons is a major approximation. We plan to remove this in a later work.
This should extend the reach of our computations to lower values of $\sqrt
S$. Another major limitation is that the hadronic approach is unlikely to
be a reasonable way to capture the physics of the chirally symmetric state
of QCD; however, that is not our concern in this work. Our main concern
is to obtain a treatment of transport theory in the chiral
symmetry broken region of QCD using a controllable, and independently
testable, hadron EFT. A step towards this is what we present here.

\section{Transport theory in the hadron phase}

\subsection{Chemical rate equations and relaxation times}

The kinetic theory underlying chemical equilibrium and freezeout is
well known \cite{Kolb:1990vq}.  In order to set up our notation and the
model approximations, we give a brief review here of the passage from
the Boltzmann to the chemical rate equations.  The Boltzmann equation
for a species $a$ in the reactive fluid can be written in the form
\beq
 D\rho_a(x,p) = C[\rho],
\eeq{boltz}
where $\rho_a(x,p)$ is the Lorentz invariant phase space density, and
$x$ and $p$ are 4-vectors for the position and momentum. We write the
Liouville operator in the form $D=p_\mu\partial^\mu$, appropriate for
rectilinear coordinates in flat space. The Lorentz vector number current
is defined as
\beq
 n^\mu_a(x) = \int d\Gamma_ap^\mu\rho_a(x,p), \qquad{\rm where}\qquad
  d\Gamma_a = \frac{g_a}{(2\pi)^3}\;\frac{d^3p}{2E},
\eeq{yield}
and $g_a$ is the phase space multiplicity factor, which counts, in
general, the dimensions of both the spin and isospin representations,
\ie, $g_a=(2S_a+1)(2I_a+1)$. In our particular application, since we
include only pseudoscalar mesons, $S_a=0$. Note that the particle number
density $n_a$ is one component of this four vector. In the following,
we will examine it in the frame in which the heat bath is at rest.

Integrating both sides of \eqn{boltz} one may then write
\beq
 \partial_\mu n_a^\mu = \int d\Gamma_a C[\rho].
\eeq{rate}
The right hand side of this chemical rate equation can be written as
\beqa
\nonumber
 \int d\Gamma_a C[\rho] &=& \int \left\{\prod_id\Gamma_i\right\}\,\left\{\prod_fd\Gamma_f\right\}
   (2\pi)^4\delta^4\left(\sum_ip_i-\sum_fp_f\right)
 \\ &&\qquad \sum_r
  \left[|M|_{r(i,f)}^2\,\left\{\prod_i\rho_i\right\}\,\left\{\prod_f(1\pm\rho_f)\right\}
        -|M|_{\bar r(f,i)}^2\,\left\{\prod_i(1\pm\rho_i)\right\}\,\left\{\prod_f\rho_f\right\}\right],
\eeqa{coll}
where the sum is over all reactions $r(i,f)$ which include the particle
$a$ among the set of initial particles $i$ and with the appropriate set of final
particles $f$. The reverse reaction $\bar r(f,i)$ interchanges the
initial and final sets. Since QCD at zero chemical potentials preserves
CP symmetry, one can equate $|M|_r^2$ and $|M|_{\bar r}^2$. Further, as
long as quantum effects like Pauli blocking or Bose enhancement can be
neglected, one can set $\rho\ll1$. As a result, $1\pm\rho\simeq1$. using
these two approximations one can write
\beq
 \int d\Gamma_a C[\rho] = \int \left\{\prod_id\Gamma_i\right\}\,\left\{\prod_fd\Gamma_f\right\}
   (2\pi)^4\delta^4\left(\sum_ip_i-\sum_fp_f\right) \sum_{r(i,f)} |M|_r^2
  \left[\left\{\prod_i\rho_i\right\}-\left\{\prod_f\rho_f\right\}\right],
\eeq{collp}
There are similar equations for the whole coupled chain of reactions. No
assumptions need to be made at this stage about whether to use Boltzmann
or quantum distributions for the distribution functions $\rho$. This is the
form developed in \cite{Kolb:1990vq}, for example.

In the specific case that is of interest here, the temperatures could be
slightly higher than the pion mass, but certainly less than twice that;
$m_\pi<T<2m_\pi$. So reactions which produce more particles in the final
state than in the initial are rare. Also, since the phase space densities
are much less than unity, collisions of three or more particles are
extremely rare. As a result, one can restrict a first investigation to
reactions involving two particles in the initial state and two in the
final state, \ie, 2-to-2 reactions. In this case, further reduction of
\eqn{collp} is possible.

The total cross section for the forward reaction $ab\to cd$ is 
\beq
 F_{ab}\sigma_r = \int d\Gamma_cd\Gamma_d
   (2\pi)^4\delta^4(p_a+p_b-p_c-p_d) |M|_r^2,
\eeq{cross}
where the Lorentz invariant definition of the flux of particles in
the initial state is $F_{ab}^2=(s-(m_a+m_b)^2)(s-(m_a-m_b)^2)$, where
$s$ is the square of the CM energy in the collision of $a$ and $b$
\cite{Cannoni:2016hro}. One can use a similar expression for the cross
section, $\sigma_{\bar r}$ of the reverse reaction, $cd\to ab$. One
may trade the squared matrix element on the right for the combination
on the left in the rate equation, if one wants to.

We will now choose to work in the frame in which the heat bath is at
rest. Our model consists of the fluid at rest in this frame, so that
the spatial components of $n_a^\mu$ vanish, and the time component is
the particle density. Then the rate equation becomes
\beq
 \frac{dn_a}{dt}=-\sum_r\langle\langle\sigma_rv_{ab}\rangle\rangle n_an_b
      +\sum_{\bar r}\langle\langle\sigma_{\bar r}v_{cd}\rangle\rangle n_cn_d.
\eeq{react}
Here we have used the notation $v_{ab}=F_{ab}/(4E_aE_b)$; a
Lorentz covariant definition of $v_{ab}$ follows from the above
definitions. Further, the double angular bracket indicates averaging
over the instantaneous densities of the initial particles for a
reaction. However, these non-equilibrium distributions are not universal,
and a general study is not of much interest. It is more useful to
examine this in the linear-response limit as the system reaches close
to equilibrium.

Accordingly, we set $n_a=\Neq_a+\delta n_a$, expand in the small
deviations from equilibrium, $\delta n_a$, \etc, and retain terms up
to linear order in these small parameters. The terms independent of
the $\delta n$s vanishes due to detailed balance.  We can then replace
averages with respect to the non-equilibrium distributions by averages
in equilibrium.  These are denoted by single angular brackets below.
To linear order the reaction rate equations become
\beq
 \frac{d\delta n_a}{dt}=
  -\sum_r\langle\sigma_rv_{ab}\rangle (\Neq_b\delta n_a+\Neq_a\delta n_b)
  +\sum_{\bar r}\langle\sigma_{\bar r}v_{cd}\rangle(\Neq_d\delta n_c+\Neq_c\delta n_d).
\eeq{req}
The linear system of equations can be represented in the form $\dot{\bf
n} = -A{\bf n}$, where $\bf n$ is a column vector whose elements are
each of the deviations of the number densities of interest from their
equilibrium values . Each element of the matrix $A$ is a quantity of the
form $\langle\sigma v\rangle n$. One sees that dimensionally this is the
inverse of a time (we have used natural units throughout).  For every
conserved quantity, one has a zero eigenvalue of $A$.  Every other
eigenvalue of $A$ is the inverse of one of the relaxation times of
the system.

The longest relaxation time tells us how fast the disturbed system
relaxes back to chemical equilibrium. Far away from equilibrium the
system can relax either faster or slower. However, as the system
approaches equilibrium, the rate of approach to thermal equilibrium
cannot be faster than the inverse of the longest relaxation time. We
relate these relaxation times to freeze out in the next subsection.

\subsection{Expansion time scale and freeze out}

If chemical freeze out occurs somewhat late in the lifetime of the
fireball, then one may approximate the fluid flow by a uniform radial
flow. It is most convenient in this case to choose radial coordinates in
the CM frame of the fireball. We can then write the Liouville operator
in these curvilinear coordinates and find that the linearized form of
the chemical rate equations become
\beq
 \frac{d{\bf n}}{dt}+\frac3t{\bf n}=-A{\bf n},
\eeq{creq}
where the components of $\bf n$ are the deviations from the equilibrium
value of each of the densities, and $A$ is the same matrix as before,
and $t$ is the elapsed time in the chosen frame. A small formal
remark: the square matrix $A$ is not symmetric. Since the vector $n$
is defined to be a column vector, and $A$ acts on it from the left,
the right eigenvectors are the ones which specify the normal modes of
the system. The eigenvectors need not be orthogonal to each other. This
equation now allows a simple dimensional argument for the freezeout time.

Each eigenvalue of $A$, $\lambda_i$, is the inverse of the relaxation
time, $\tau_i=1/\lambda_i$, of one of the eigenmodes of the fluid when
there is no flow. Every conservation law gives an exact zero eigenvalue;
these may be disregarded for the analysis of freeze out. If the other
$\tau_i\ll t/3$, then the any deviation from equilibrium dies away much
faster than the expansion rate, and the fluid may be considered to be
in chemical equilibrium as it expands. If one of the $\tau_i>t/3$,
then the corresponding eigenmode of $A$ will not be able to relax back
to equilibrium, and that mode may be considered to have frozen out. The
subsequent evolution of the fluid involves the frozen mode(s) as well
as modes which may remain in equilibrium. If there is a big hierarchy
between the chemical relaxation times, $\tau_i$, then it may be possible
to use sequential freeze out scenarios, which are common in cosmology
\cite {Kolb:1990vq}, and have been proposed in heavy ion collisions
\cite{variants}.

\bef[bth]
 \begin{center}
 \includegraphics[scale=0.5]{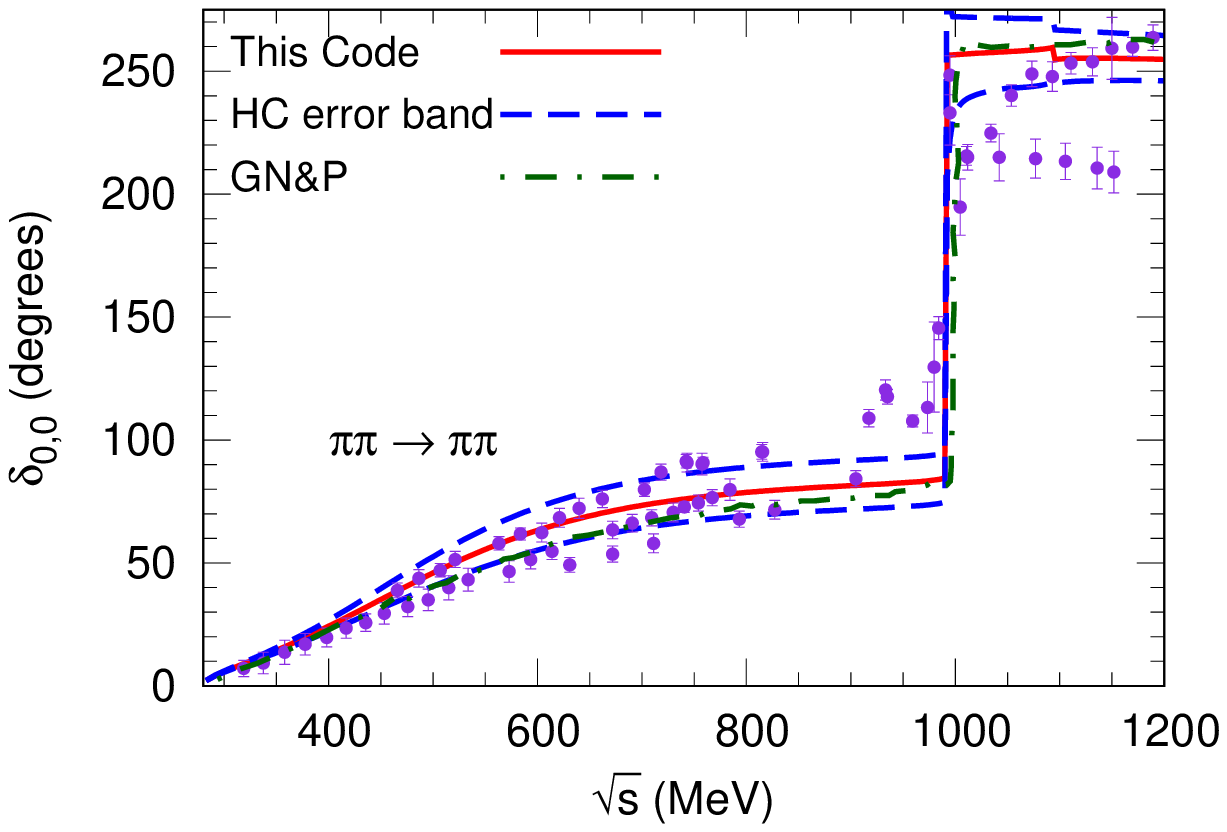}
 \includegraphics[scale=0.5]{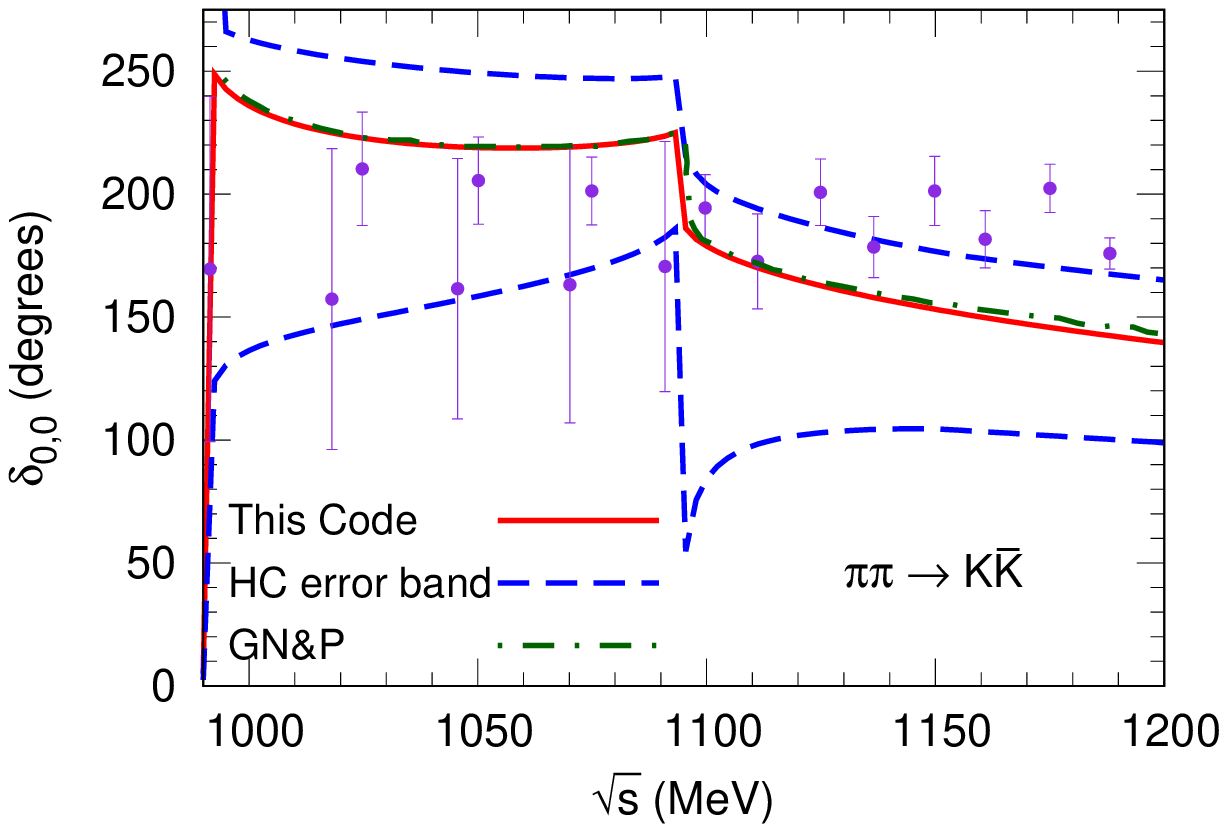}	%
 \includegraphics[scale=0.5]{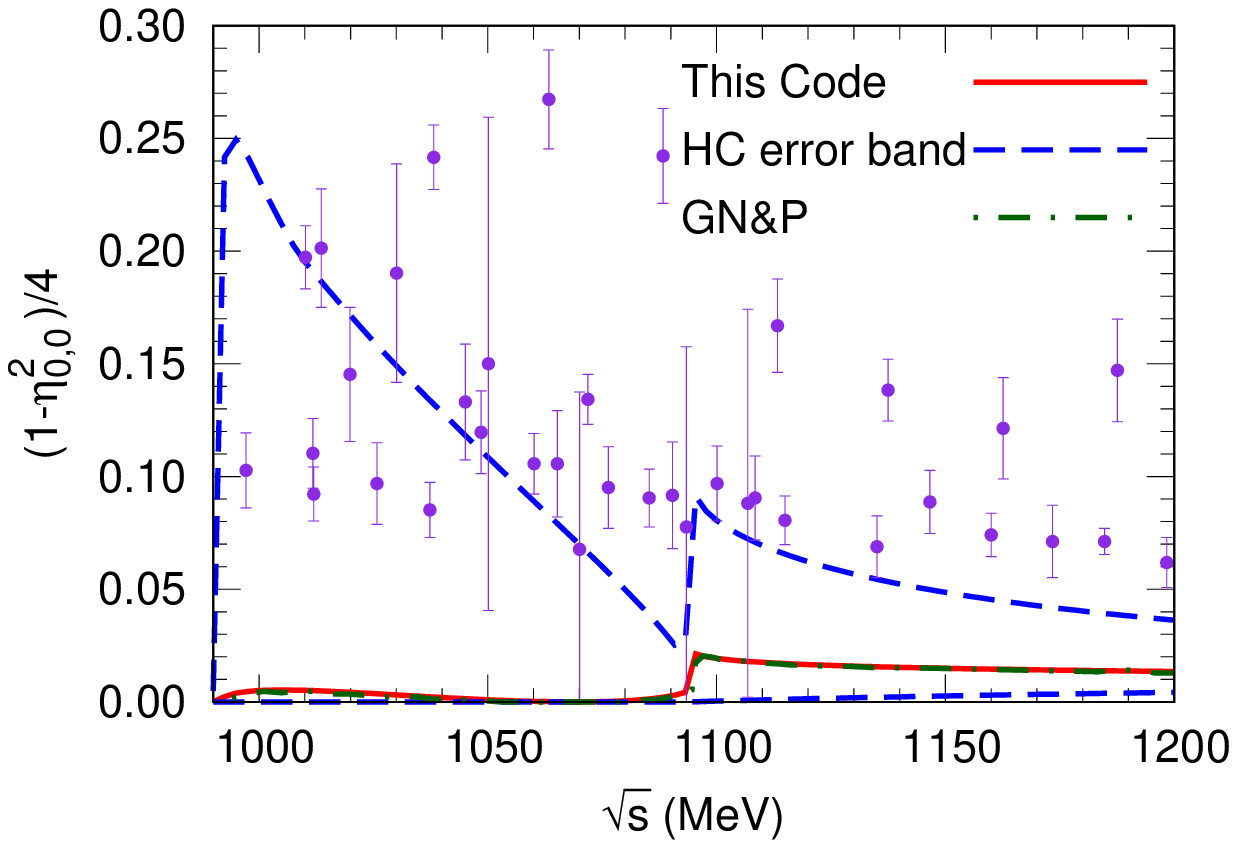}
 \includegraphics[scale=0.5]{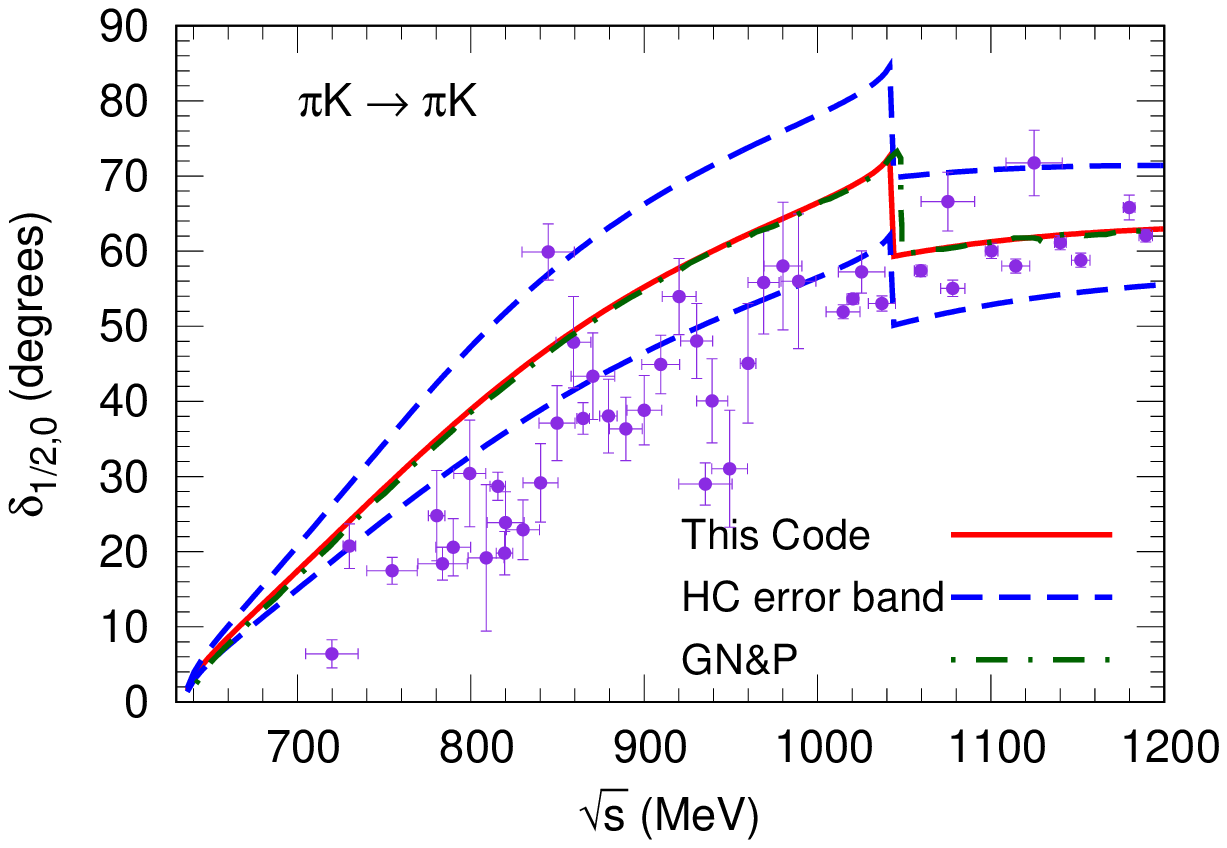}	%
 \includegraphics[scale=0.5]{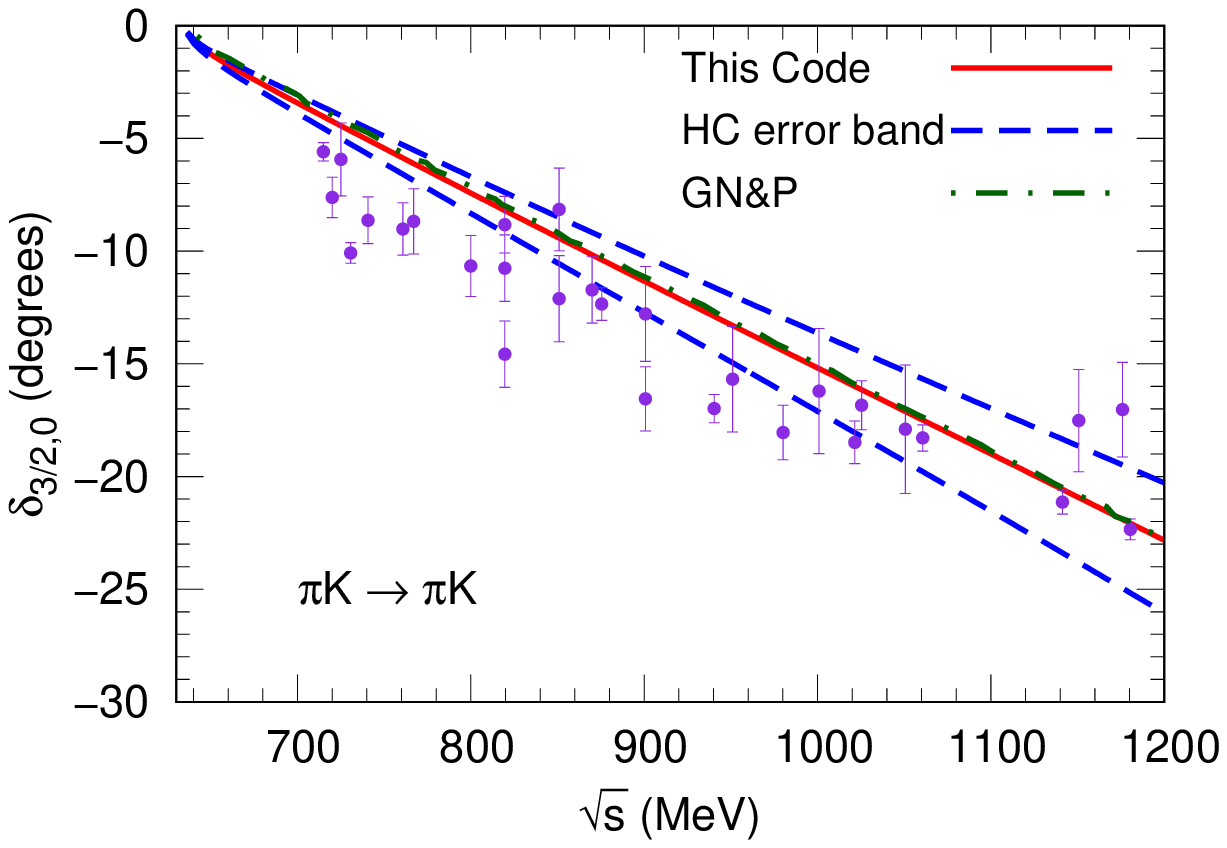}
 \includegraphics[scale=0.5]{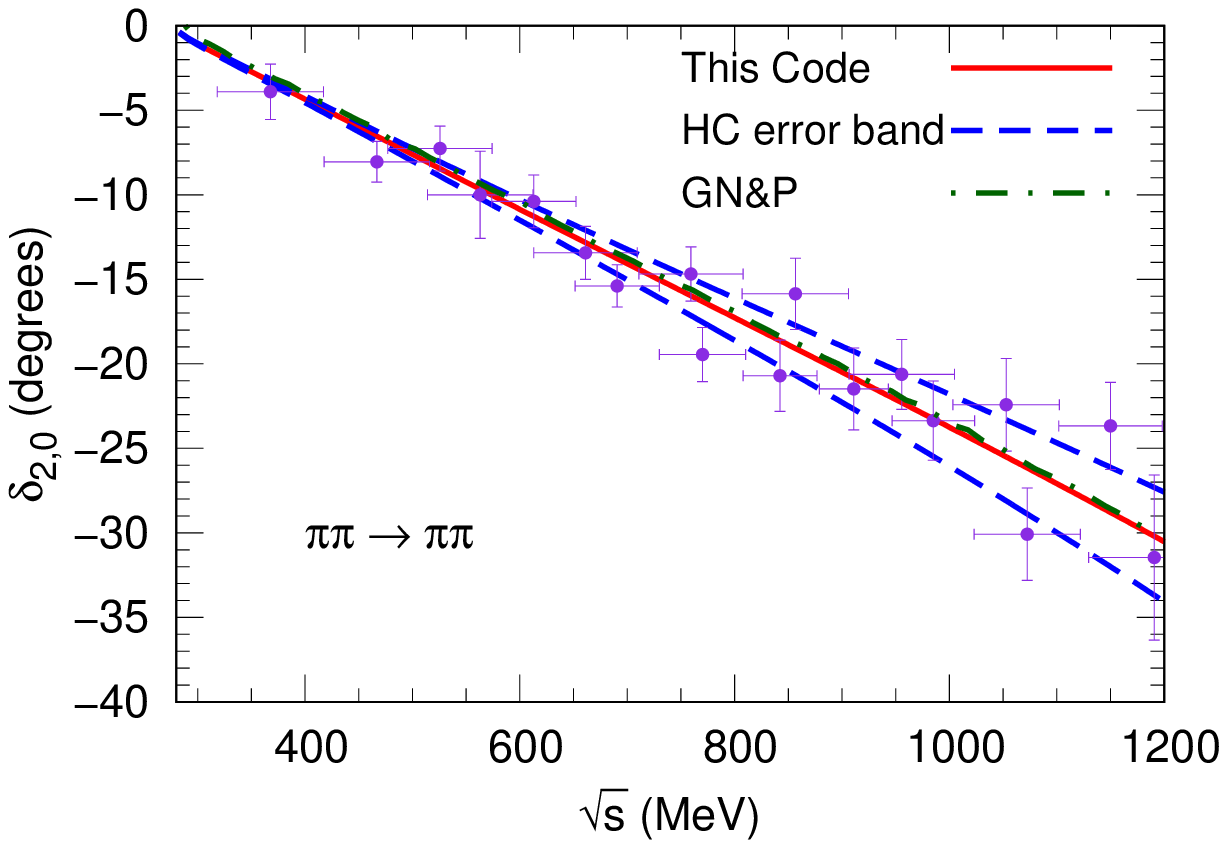}		%
 \includegraphics[scale=0.5]{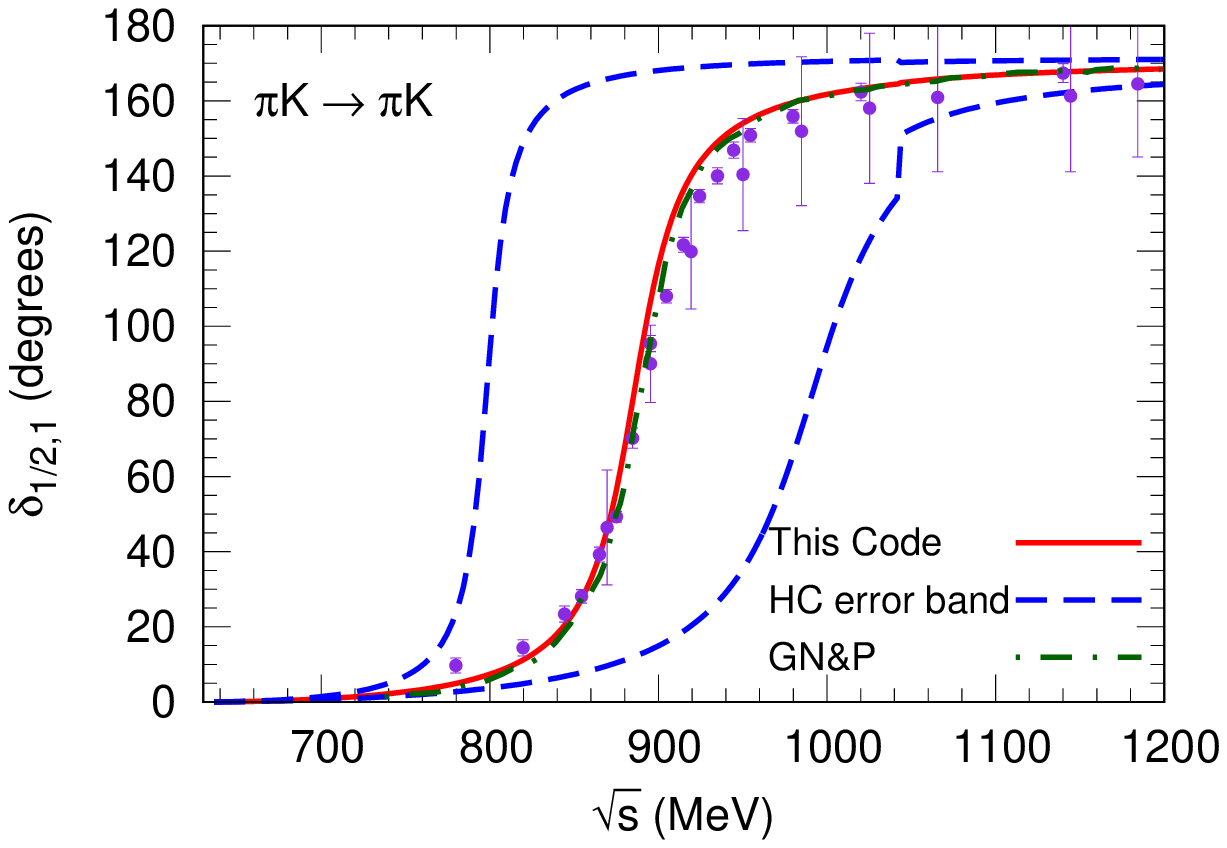}
 \includegraphics[scale=0.5]{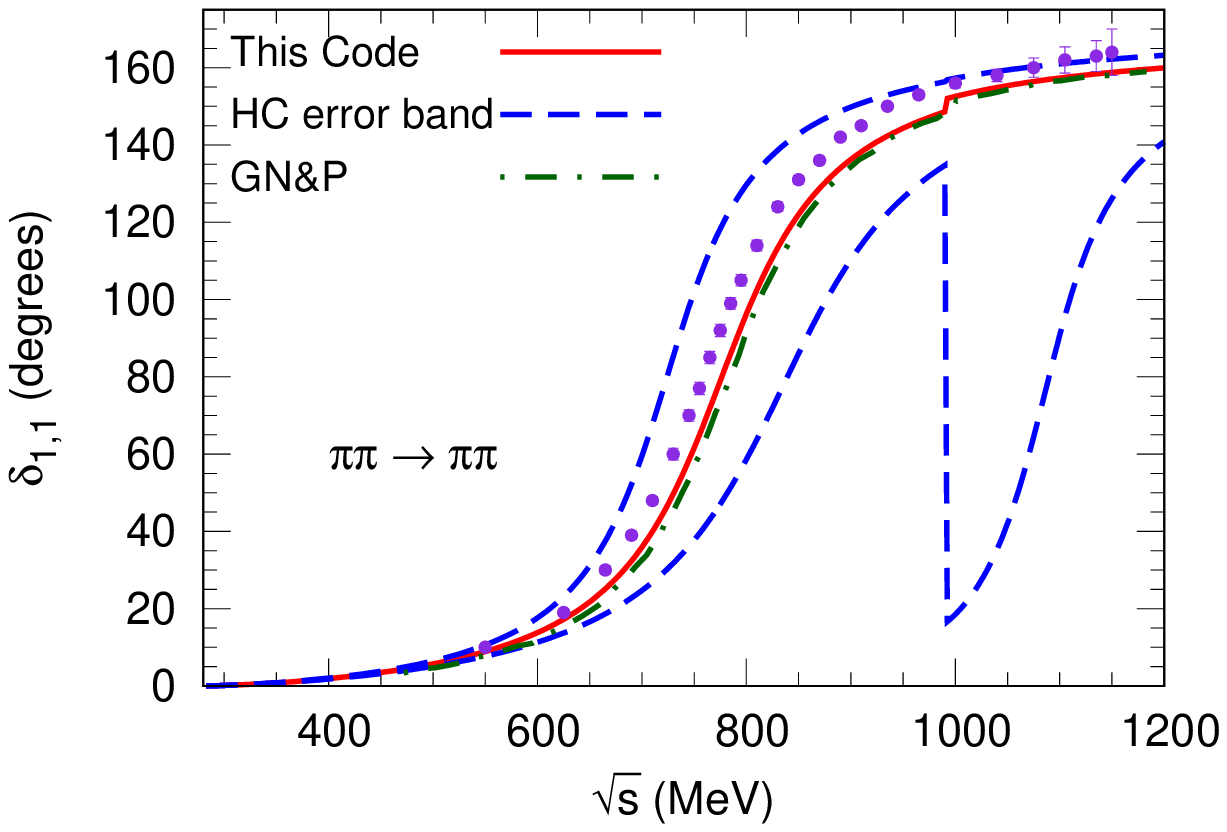}
 \end{center}
\caption{Scattering phase shifts, $\delta_{I,J}$, and inelasticity
 $\eta_{I,J}$ in the scattering of pions and kaons for isospin, $I$, and
 angular momentum, $J$, channels, shown as a function of the CM energy of
 the colliding mesons ($\sqrt s$). The dash-dotted lines (labelled GN\&P)
 shows the results reported in \cite{GomezNicola:2001as}, the full line
 shows the central values of our results, and the dashed lines denote
 the minimum and maximum when varying the input parameters among the corners
 of the error hypercube.}
\eef{phsshift}

We note that at very large times, $t$, the formal solution of the
above equation may still allow $\tau_i\ll t$. However, the system
will have diluted to the point that a hydrodynamic flow is no longer
feasible. Even if the $\tau_i$ are so small that radial flow is not a very
good approximation at $t\approx\tau_i$, a dimensional argument shows that
hydrodynamic expansion of the fireball would change the factor $3/t$
to $\xi/t$ where the dimensionless number $\xi$ depends on the details
of the flow.  With this change the arguments given above would continue
to be applicable.

\subsection{Hadron cross sections from chiral perturbation theory}

As discussed earlier, in this paper we report an investigation of a
model hadron fluid made of the pseudo-Goldstone bosons of chiral symmetry
breaking, namely the lowest SU(3) flavour octet of pseudoscalar bosons.
The mutual interactions of these mesons are completely constrained by
chiral symmetry \cite{Weinberg:1978kz, Gasser:1983yg, Amoros:2000mc}, and
their reaction cross sections have been computed in chiral perturbation
theory. We use the unitarized amplitudes which were presented in
\cite{GomezNicola:2001as}. These depend on the meson masses, $f_\pi$, and
8 other LECs ($L_1$, $L_2\cdots$, $L_8$) which appear in the Lagrangian
of chiral perturbation theory to order $p^4$.

Discussions of the extraction of the LECs from hadron
observables can be found in \cite{Gasser:1983yg, Amoros:2000mc}.  $L_1$,
$L_2$, and $L_3$ can be determined through $\pi\pi$ scattering in the
$J=2$ channel and from the weak decays of $K$. Constraints on $L_1$,
$L_3$, $L_4$, and $L_6$ come from Zweig's rule. $L_5$ may be determined
by the $f_\pi$ to $f_{\scriptscriptstyle K}$ ratio.  $L_4$ and $L_6$
are also obtained from the $\pi$ scalar and charge radii.  $L_5$, $L_7$
and $L_8$ are constrained by the Gell-Mann-Okubo mass formulae. Currently
error bands in individual constants range from 25--40\%. In this work
we shall exhibit the uncertainties in predictions due to the current
range of uncertainties in these constants.

Two implementations of unitarized amplitudes from chiral perturbation
theory are discussed in \cite{GomezNicola:2001as}. In one the dimension
of the T-matrix is taken to be equal to the number of channels which are
open, so that the dimension changes at every mass threshold. This strictly
implements unitarity at each energy, but may result in a discontinuity
of the amplitude at thresholds where new channels open up. The other
is to use a constant dimension for the T-matrix, equal to the highest
dimension required until about 2 GeV. This may result in a loss of
unitarity in the vicinity of a threshold, but preserved continuity.
The difference between the two methods for phase shifts was shown to
be minor. In our implementation we chose the first approach, so that
unitarity is always strictly implemented. 

In \fgn{phsshift} we compare our numerical implementation with that of
\cite{GomezNicola:2001as}.  In the figures we have also given the theory
errors on the phase shifts induced by the errors in the input low energy
constants. These are estimated by taking the maximum and minimum values
of the phase shifts as we vary the parameters along the corners of a
hypercube whose center is at the central values of the parameters, and
whose faces are one sigma away.  We note that keeping track of theory
errors is important, as one sees in \fgn{phsshift}. We also note the
spread in measured values of scattering amplitudes from several different
experiments.

\bef[bth]
 \begin{center}
 \includegraphics[scale=0.5]{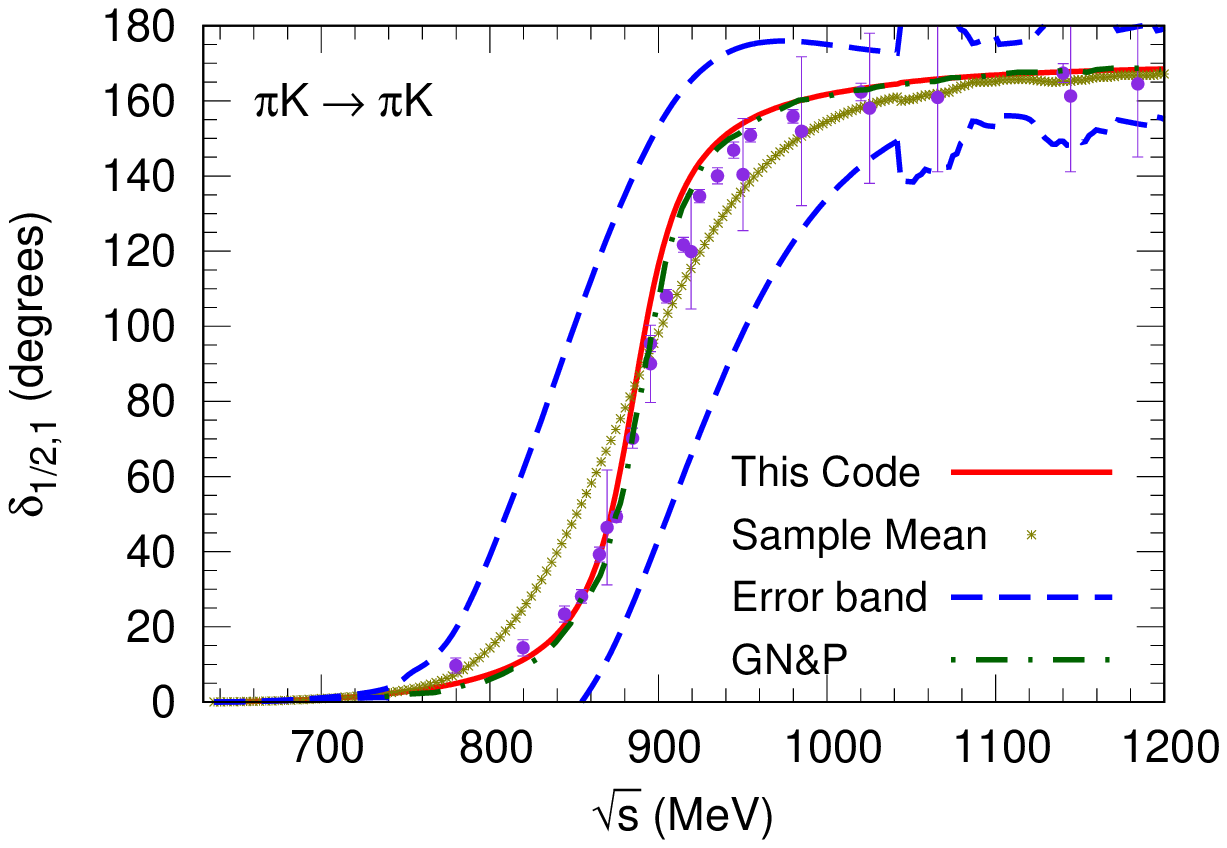}
 \includegraphics[scale=0.5]{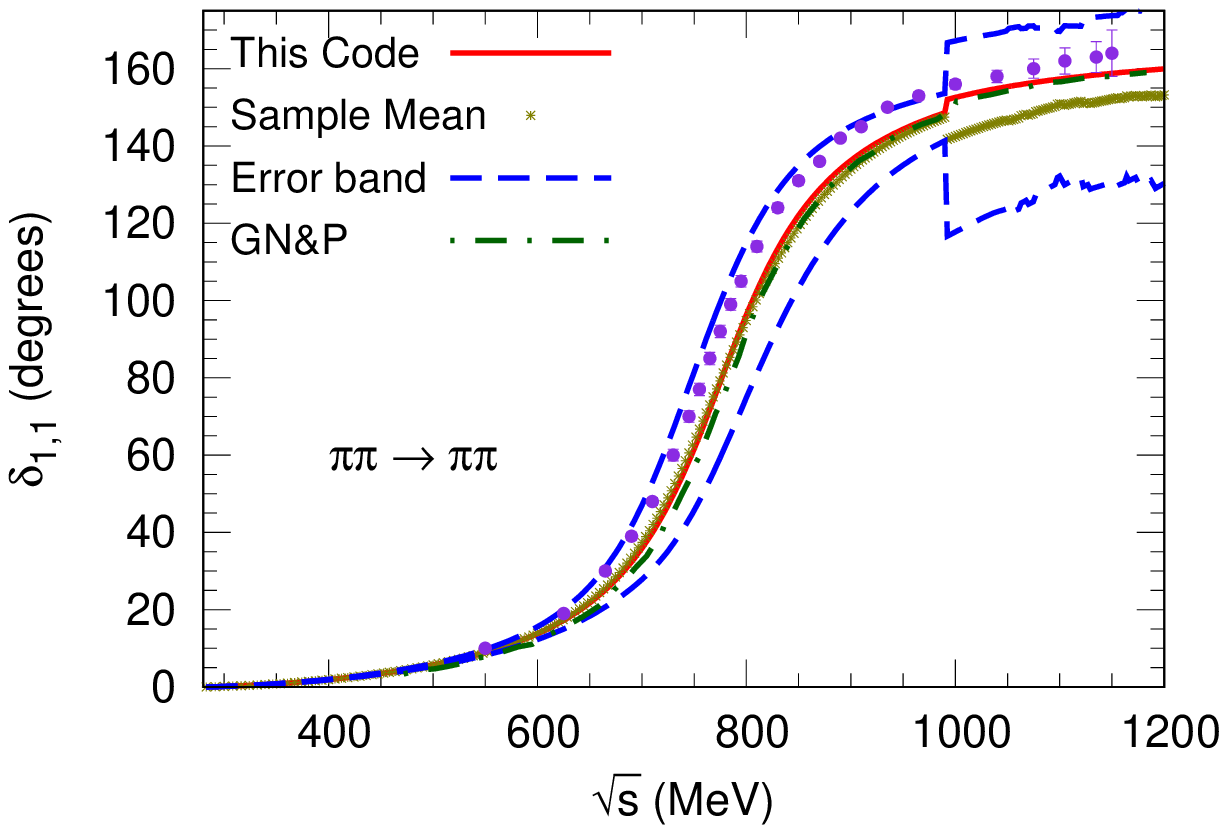}
 \end{center}
\caption{Alternative estimates of the theory uncertainty in scattering
 phase shifts, $\delta_{I,J}$, in two channels with resonances. The
 solid curve gives the values when all the LECs are fixed at their best
 values. Also shown are Monte Carlo estimates using 500 samples of the
 LECs, when each is sampled independently by a Gaussian centered at the
 best value and width equal to the quoted error (used earlier to fix the
 corners of the uncertainty hypercube). The crosses are the mean value,
 and the dashed lines are the one sigma errors from this Monte Carlo.}
\eef{gauss}

However, some of the errors estimated by moving along the corners of the
hypercube are quite extreme, and are dominated by a single parameter. The
phase shift $\delta_{1,1}$ is one such. The magnitude of the discontinuity
of the error at the threshold for $K$ pair production is probably an
artifact. Because of this we investigate a more natural, but more
expensive, estimator. We draw samples of the LECs from independent
Gaussian distributions for each parameter with a mean equal to the
central value and width equal to the quoted error. Two examples of the
error bands obtained by this procedure are shown in \fgn{gauss}. Note
that the effect of a single outlier is no longer as important. We will use
this method in the rest of the paper.

Note also that $\delta_{11}$ and $\delta_{1/2,1}$ show very clearly the
presence of the $\rho$ and $K^*$ mesons. A closer investigation shows
that the interacting gas of pseudoscalar mesons generates the full SU(3)
octet of vector mesons. We also checked that the interacting system
generates all scalar mesons with masses less than 2 GeV. There are no
tensor mesons in this mass range, so we the interacting system of SU(3)
octet of pseudoscalars generates the hadron resonance gas of all SU(3)
octets with masses less than 2 GeV. We use these amplitudes to compute
the elements of the matrix $A$ defined after \eqn{creq}.

We note that an earlier work along these lines \cite{Prakash:1993bt}
also tried to use chiral perturbation theory.  However, that work predated
\cite{GomezNicola:2001as}, as a result of which the resonance region was
then not captured by this fundamental approach.  So, parametrizations were
used in \cite{Prakash:1993bt} to describe amplitudes in the resonance
region. We are able to avoid this due to advances in QCD, which allows us
to proceed entirely by using a state-of-the-art effective field theory (EFT)
with a small number of low energy constants, namely chiral perturbation
theory. This allows us to use a small number of parameters which are
determined by measurements of other hadron properties, and to write unitary
amplitudes for the transport theory without any ad-hoc UV cutoff.

\section{Results}

In this paper we neglect the differences in masses due to SU(2) isospin
breaking, and treat all pions as having a mass equal to the mean mass
of the isotriplet.  We also take all the kaons to have exactly equal
mass, that of the mean mass of each isodoublet. The matrix elements then
distinguish four classes of mesons: pions, whose number density we denote
by $\npi$, kaons (with strangeness of $-1$) having number density $\nk$,
anti-kaons (strangeness of $+1$) with number density $\nkb$, and $\eta$
with number density $\neta$. The linearised reaction matrix $A$ is then
a $4\times4$ matrix.

Collapsing each isospin multiplet into one species means that we cannot
examine thermalization times associated with isospin fluctuations.
This is an interesting issue which could be addressed by a fundamental
approach such as the one we take. However, the main interest in this
question is due to isospin fluctuations in the baryon sector, so we
defer this question for later.

The computation of the reaction rate matrix $A$ requires $\Neq(m,T)$. We
use the expansion of the number density from the Bose distribution
\beq
 \Neq(m,T) = \frac{m^2T}{2\pi^2}\sum_{n=1}^\infty\frac1n\,
      K_2\left(n\,\frac mT\right),
\eeq{bosexp}
where $K_2$ is the modified Bessel function of order 2 which decays
exponentially as a function of its argument. For temperatures in the
range up to 160 MeV or so, ten terms of the series are sufficient to
get the number density accurate to five places of decimals for the pion.
For the remaining mesons the leading term suffices. This corresponds to
the classical Boltzmann gas approximation.

The unitarized amplitudes require a partial wave expansion of the
expressions obtained from chiral perturbation theory, and, up to the order
at which the amplitudes are available, it is sufficient
to keep terms only up to $J=2$. It is most convenient to
do this expansion numerically.  Our codes used standard LINPACK routines
for unitarization and inversion of matrices, and QUADPACK routines
for integrals.  Since our matrices are extremely small ($3\times3$ at
most), we also experimented with simpler high accuracy in-line code for
eigenvalues and inverses; these improved run times slightly without
compromising accuracy. For the hadronic matrix elements, we found
unitarity of the S-matrix in all channels up to machine precision for
center of mass energies of a little over 2 GeV. Computations of $\langle
\sigma v\rangle$ used Gauss-Laguerre routines for the integration
over momenta, and Gauss-Legendre routines for angular integrations. We
checked that numerical rounding and truncation errors lie well below
the theoretical uncertainties due to the input hadron parameters. The
comparisons shown in \fgn{phsshift} are part of the check of our codes.

\bef[tbh]
 \begin{center}
  \includegraphics[scale=0.8]{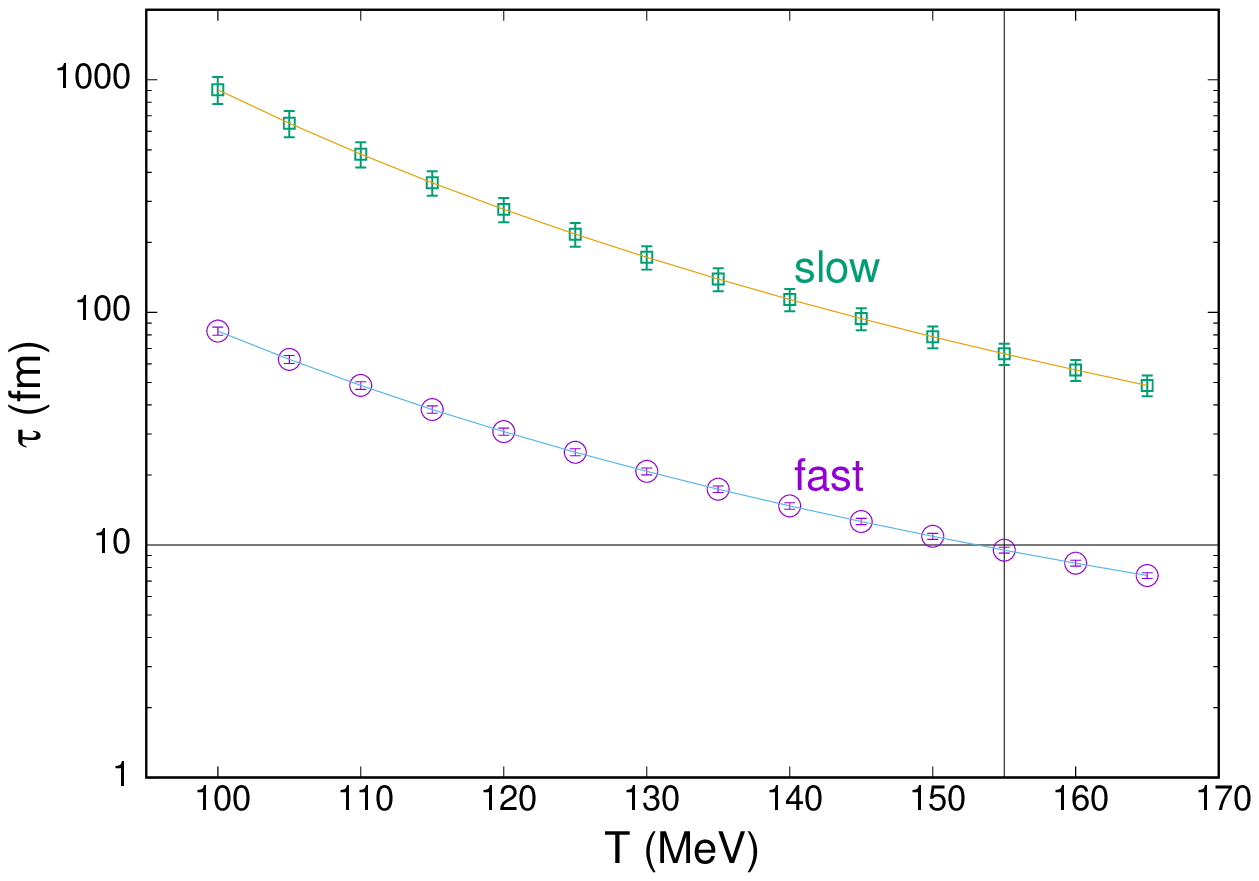}
  \includegraphics[scale=0.8]{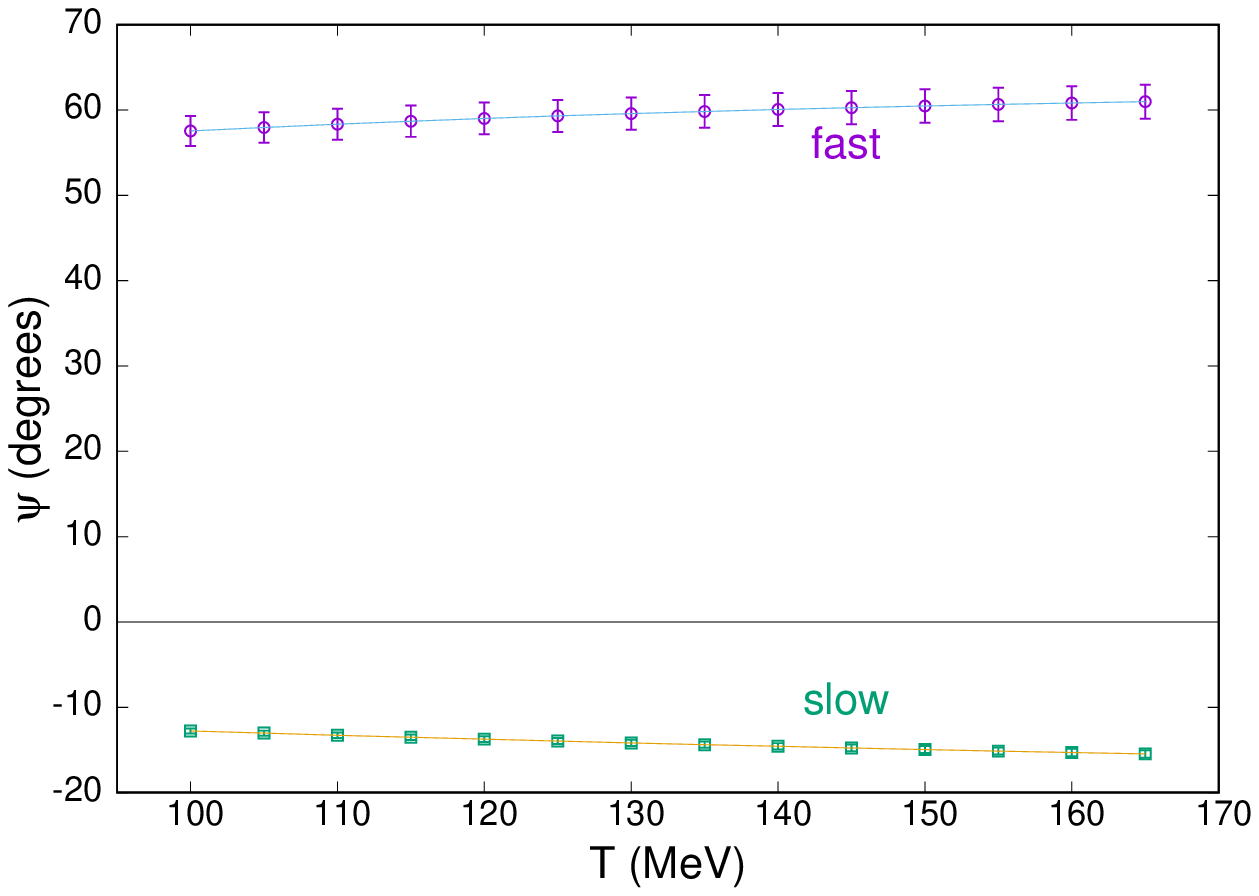}
 \end{center}
 \caption{The relaxation times, $\tau$, for the two normal modes of the
  linearized chemical rate equations are shown in the first figure as a
  function of the temperature. The eigenmodes are shown in the second
  figure; $\psi$ is the angle the eigenvector makes with the pion direction.
  In both cases, the error bands shown are obtained from a Monte Carlo
  sampling of parameters, as described earlier.}
\eef{results}

There are two conserved quantities--- net strangeness, and total particle
number (since we used only 2-to-2 reactions). As a result there are two
eigenvalues of $A$ which vanish. It is simple to actually parametrize the
number densities so that this is taken account of. We can write $\npi=
\Neq_\pi+h_\pi$ and $\neta=\Neq_\eta+h_\eta$. Then one has 
$\nk=\Neq_{\scriptscriptstyle K}-(h_\pi+h_\eta)/2$ and
$\nkb=\Neq_{\overline{\scriptscriptstyle K}}-(h_\pi+h_\eta)/2$. Substituting
these parametrizations in \eqn{creq}, one can eliminate two of the equations
to obtain the reduced set of equations
\beq
 \frac{d{\bf h}}{dt}+\frac3t{\bf h}=-C{\bf h},
\eeq{rcreq}
where $\bf h$ is a two dimensional column vector whose components are
$h_\pi$ and $h_\eta$ respectively, and $C$ is a reduced matrix obtained
from $A$ by eliminating the equations for $\nk$ and $\nkb$ using the
conservation laws. The eigenvalues of $C$ are the inverses of the relaxation
times of the system. The two eigenvectors can be specified by the angle,
$\psi$, that they make with the $h_\pi$ axes.

The results of our computations are shown in \fgn{results}.  At a
temperature of about 150 MeV, the fast mode has a relaxation time of
around 10 fm. This increases to about a 100 fm at a temperature of 100
MeV. The slow mode is an order of magnitude slower, with a relaxation
time of around 100 fm at 150 MeV. The slow mode is dominantly of pions
relaxing towards equilibrium; the eigenvector makes an angle of 10 to
15 degrees with the pion direction. The fast mode is dominated by the
relaxation of the $\eta$, since it makes an angle of about 30 degrees
with the $\eta$ direction.

Relaxation times of 100 fm may seem unnatural in a hadron gas, where
typical time scales are expected to be around 1 fm. However, one may
recall that $1/\tau\simeq\sigma\Neq(m_\pi,T)$, where $\sigma$ is an
average over the  hadron cross sections and $\Neq(m_\pi,T)$ is given by
\eqn{bosexp}. The fact that the eigenvector corresponding to the slow
mode is almost fully aligned with the pion direction shows that this
assumption is fairly accurate. In a theory of pions, one would expect
$\sigma\simeq m_\pi^2/f_\pi^4$.  In fact, one finds that the dimensionless
ratio $\Pi_s=\tau_s(T)\Neq(m_\pi,T)m_\pi^2/f_\pi^4$ (here $\tau_s$ is the
relaxation time of the slow mode) is of order unity, and varies from 4 to
8 over the range of temperature shown in \fgn{results}. This is a check
that the very long relaxation time is natural, and that its magnitude is
what should be expected {\sl a priori\/} from chiral perturbation theory.

With the relaxation time of the fast mode, $\tau_f$, one can similarly
compute the dimensionless ratio $\Pi_f = \tau_f(T) \Neq(m_\eta,T) m_\pi^2
/f_\pi^4$. Again, this turns out to be of order unity, varying from 0.5
to 1 over the same range of temperature. The use of the number
density for $\eta$ in this case is suggested by the large component of
the fast eigenmode in the $\eta$ direction. Both the fast and slow modes
turn out to be natural. The unexpected size of the relaxation times is
due essentially to the fact that the equilibrium densities of the mesons
at these temperatures are small.

The interactions of Goldstone bosons are strongly constrained, and involve
derivative interactions which are forced by symmetry.  The fact that quark
masses are non-zero allow non-derivative couplings.  A signal of this is
that the mode dominated by the highest mass pseudo-Goldstone boson, the
$\eta$, has lower relaxation time.  We can check this also by dropping
the $\eta$ meson from the computation. Removing it should decrease the
net cross sections, and push up the relaxation time. After removing
the two conservation laws using the equations for $\nk$ and $\nkb$,
and writing the equation for the pion in terms of $h_\pi$, its deviation
from equilibrium, one can write the linearized chemical rate equation as
\beq
   \frac{dh_\pi}{dt}+\frac3th_\pi = -\frac1{\tau'} h_\pi,
\eeq{oneeq}
Where the relaxation time, $\tau'$, of the gas without the $\eta$ is
computed in terms of the remaining rates. The result is about an order
of magnitude larger than $\tau_s$. The movement is in the direction that
we expected from the argument based on chiral perturbation theory.

\section{Conclusions}

We examined the chemical relaxation time in a gas of the SU(3) octet of
pseudoscalar mesons in the linear response approximation. The reaction
amplitudes in the gas were described by the best current results in
chiral perturbation theory available to date \cite{GomezNicola:2001as}.
These interactions generated the full SU(3) octet of vector mesons,
and all the scalars up to 2 GeV in mass. The amplitudes were completely
parametrized by three masses, one decay constant, and eight low energy
constants of chiral perturbation theory, all of which were known from
experimental observables.  We found that the slowest relaxation time,
$\tau_s$, which controls the rate of approach to equilibrium, is about
100 fm. The second relaxation time, $\tau_f$, was found to be of the
order of 10 fm.  We argued that these long relaxation times are natural
for a gas of pseudo-Goldstone bosons.  This then implies that such a gas
cannot equilibrate chemically within the lifetime of a fireball produced
in heavy-ion collisions.

However, this leads to an interesting conjecture. While this study has
nothing to say about chemical relaxation times in the chiral symmetric
phase of a strongly interacting system, one may assume, as is common,
that hot QCD matter has short relaxation times. Then it is entirely
possible for this system to come to equilibrium in the early history of
the fireball. Our results show that as the system passes through the QCD
chiral cross over, $T_{co}$, and the degrees of freedom change to the
pseudo-Goldstone boson of the broken chiral symmetry, it is not possible
to maintain chemical equilibrium. This gives a straightforward mechanism
for $T_{fo}$ to occur very close to $T_{co}$.  Mild mismatches between
the two temperatures are entirely possible due to the fact that a cross
over is blurry.  This explanation is pleasant because the very mechanism
that causes the cross over also causes chemical freezeout.

It should be noted that there are escape routes to this argument.
The first is to note that this computation does not include baryons,
whereas significant fraction of heavy-ion collision events contain at
least a few baryons in the fireball. We are currently investigating
a mixture of baryons and mesons, and will report on this aspect later.
The next route is technical. The formalism we use is based on a reduction
of the Boltzmann transport equations which assumes that the phase
space density of particles is significantly less than unity (in natural
units). If such deviations from the results of a hadron resonance gas
model can be observed, then further investigations will be required.
Such an eventuality has been suggested, but has not yet been observed
in heavy ion collisions.


\begin{thebibliography}{99}
\bibitem{Andronic:2017pug}
 A.~Andronic, P.~Braun-Munzinger, K.~Redlich and J.~Stachel,
 Nature \textbf{561} (2018) no.7723, 321-330
 [arXiv:1710.09425 [nucl-th]].
\bibitem{variants}
 J.~Cleymans, M.~Marais and E.~Suhonen,
 Phys. Rev. C \textbf{56}, 2747-2751 (1997)
 [arXiv:nucl-th/9705014 [nucl-th]];\\
 A.~Majumder and V.~Koch,
 Phys. Rev. C \textbf{68}, 044903 (2003)
 [arXiv:nucl-th/0305047 [nucl-th]];\\
 V.~Begun, L.~Ferroni, M.~I.~Gorenstein, M.~Gazdzicki and F.~Becattini,
 J. Phys. G \textbf{32}, 1003-1020 (2006)
 [arXiv:nucl-th/0512070 [nucl-th]];\\
 T.~Csorgo, R.~Vertesi and J.~Sziklai,
 Phys. Rev. Lett. \textbf{105}, 182301 (2010)
 [arXiv:0912.5526 [nucl-ex]];\\
 J.~Cleymans and D.~Worku,
 Mod. Phys. Lett. A \textbf{26}, 1197-1209 (2011)
 [arXiv:1103.1463 [hep-ph]];\\
 R.~Singh, L.~Kumar, P.~K.~Netrakanti and B.~Mohanty,
 Adv. High Energy Phys. \textbf{2013}, 761474 (2013)
 [arXiv:1304.2969 [nucl-ex]];\\
 S.~Chatterjee, R.~Godbole and S.~Gupta,
 Phys. Lett. B \textbf{727}, 554-557 (2013)
 [arXiv:1306.2006 [nucl-th]];\\
 V.~Begun, W.~Florkowski and M.~Rybczynski,
 Phys. Rev. C \textbf{90}, no.5, 054912 (2014)
 [arXiv:1405.7252 [hep-ph]];\\
 D.~Oliinychenko, K.~Bugaev, V.~Sagun, A.~Ivanytskyi, I.~Yakimenko, E.~Nikonov, A.~Taranenko and G.~Zinovjev,
 [arXiv:1611.07349 [nucl-th]];\\
 V.~Vovchenko, M.~I.~Gorenstein and H.~Stoecker,
 Phys. Rev. C \textbf{98}, no.6, 064909 (2018)
 [arXiv:1805.01402 [nucl-th]];\\
 J.~Cleymans, B.~Hippolyte, M.~W.~Paradza and N.~Sharma,
 Int. J. Mod. Phys. E \textbf{28}, no.09, 1940002 (2019)\\
 S.~Gupta, D.~Mallick, D.~K.~Mishra, B.~Mohanty and N.~Xu,
 [arXiv:2004.04681 [hep-ph]];\\
 S.~Bhattacharyya, D.~Biswas, S.~K.~Ghosh, R.~Ray and P.~Singha,
 Phys. Rev. D \textbf{101}, no.5, 054002 (2020)
 [arXiv:1911.04828 [hep-ph]].
\bibitem{Aoki:2009sc}
 Y.~Aoki, S.~Borsanyi, S.~Durr, Z.~Fodor, S.~D.~Katz, S.~Krieg and K.~K.~Szabo,
 JHEP \textbf{06}, 088 (2009)
 [arXiv:0903.4155 [hep-lat]].
\bibitem{Bazavov:2018mes}
 A.~Bazavov \textit{et al.} [HotQCD],
 Phys. Lett. B \textbf{795}, 15-21 (2019)
 [arXiv:1812.08235 [hep-lat]].
\bibitem{Prakash:1993bt}
 M.~Prakash, M.~Prakash, R.~Venugopalan and G.~Welke,
 Phys. Rept. \textbf{227}, 321-366 (1993)
\bibitem{Goldstone:1962es}
 J.~Goldstone, A.~Salam and S.~Weinberg,
 Phys. Rev. \textbf{127}, 965-970 (1962)
\bibitem{Weinberg:1978kz}
 S.~Weinberg,
 Physica A \textbf{96}, no.1-2, 327-340 (1979)
\bibitem{Gasser:1983yg}
 J.~Gasser and H.~Leutwyler,
 Annals Phys. \textbf{158}, 142 (1984)
\bibitem{Amoros:2000mc}
 G.~Amoros, J.~Bijnens and P.~Talavera,
 Nucl. Phys. B \textbf{585}, 293-352 (2000)
 [arXiv:hep-ph/0003258 [hep-ph]].
\bibitem{GomezNicola:2001as} 
 A.~Gomez Nicola and J.~R.~Pelaez,
 Phys.\ Rev.\ D {\bf 65}, 054009 (2002)
 [hep-ph/0109056].
\bibitem{Lin:2004en} 
  Z.~W.~Lin, C.~M.~Ko, B.~A.~Li, B.~Zhang and S.~Pal,
  Phys.\ Rev.\ C {\bf 72}, 064901 (2005)
  [nucl-th/0411110];
  Z.~W.~Lin,
  Indian J.\ Phys.\  {\bf 85}, 837 (2011).
\bibitem{Bleicher:1999xi} 
  M.~Bleicher {\it et al.},
  J.\ Phys.\ G {\bf 25}, 1859 (1999)
  [hep-ph/9909407];
  H.~Petersen, M.~Bleicher, S.~A.~Bass and H.~Stocker,
  arXiv:0805.0567 [hep-ph].
\bibitem{Tanabashi:2018oca} 
  M.~Tanabashi {\it et al.} [Particle Data Group],
  Phys.\ Rev.\ D {\bf 98}, no. 3, 030001 (2018).
\bibitem{qmech}
  L.\ D.\ Landau and E.\ M.\ Lifshitz, Vol 3, ``Quantum Mechanics'', Elsevier (2005);\\
  J.\ J.\ Sakurai, ``Modern Quantum Mechanics'', Addison-Wesley-Longman (1999).
\bibitem{Hoogland:1974cv}
  W.~Hoogland {\it et al.},
  Nucl.\ Phys.\ B {\bf 69} (1974) 266.
\bibitem{Marzano:1977yc}
  F.~Marzano {\it et al.} [Amsterdam-CERN-Nijmegen-Oxford Collaboration],
  Nucl.\ Phys.\ B {\bf 123} (1977) 203.
\bibitem{Kolb:1990vq}
  E.~W.~Kolb and M.~S.~Turner,
  Front. Phys. \textbf{69}, 1-547 (1990)
\bibitem{Cannoni:2016hro}
  M.~Cannoni,
  Int. J. Mod. Phys. A \textbf{32}, no.02n03, 1730002 (2017)
  [arXiv:1605.00569 [hep-ph]].
\end{thebibliography}
\end{document}